\newif\ifpaper
\papertrue

\ifpaper
\documentclass[compsoc,conference,a4paper,10pt,times]{IEEEtran}
\else
\documentclass[12pt,twoside]{book}
\fi

\usepackage{cite}
\usepackage{amsmath,amssymb,amsfonts}
\usepackage{graphicx}
\usepackage{textcomp}
\usepackage{xcolor}
\usepackage{url}  
\usepackage{tabularx}
\usepackage{etoolbox}  
\usepackage{adjustbox}
\usepackage{tikz}
\usetikzlibrary{arrows}

\usepackage{hyperref}
\hypersetup{
    colorlinks=true,
    allcolors=.,
    urlcolor=blue,
}

\newtoggle{paper}
\ifpaper
\toggletrue{paper}
\else
\togglefalse{paper}
\fi

\iftoggle{paper}{  
  \pagestyle{plain}
} {  
  \usepackage[T1]{fontenc}
  \usepackage[utf8x]{inputenc}
  \usepackage{mathptmx}

  \usepackage[a4paper, inner=3cm, outer=2cm, top=2cm, bottom=2cm]{geometry}
  
  \usepackage[main=english,hebrew]{babel}
  
  \usepackage{ifoddpage}
  
  \usepackage[nottoc,numbib]{tocbibind}
  
  \usepackage{fancyhdr}
  \newcommand{\setfancyheaderfooter}{%
    \pagestyle{fancy}
    \renewcommand{\chaptermark}[1]{%
      \markboth{\MakeUppercase{%
      \chaptername}\ \thechapter.%
      \ ##1}{}}
    \fancyhead[LE,RO]{\textbf{\textsl{\thepage}}}  
    \fancyhead[LO,RE]{\textsl{\leftmark}}  
    \fancyfoot{}
  }
}

\iftoggle{paper}{
  \newcommand{\level}[1]{\section{#1}}
  \newcommand{\sublevel}[1]{\subsection{#1}}
  \newcommand{\subsublevel}[1]{\subsubsection{#1}}
} {
  \newcommand{\level}[1]{\chapter{#1}}
  \newcommand{\sublevel}[1]{\section{#1}}
  \newcommand{\subsublevel}[1]{\subsection{#1}}
}

\makeatletter
\newcommand{\enableopenany}{%
  \@openrightfalse%
}

\newcommand{\disableopenany}{%
  \@openrighttrue%
}
\makeatother

\iftoggle{paper}{
}{
  \newcommand{\newchapterevenpage}{%
    \enableopenany
    \chapter*{}
    \checkoddpage
    \ifoddpage
      \newpage
    \else
    \fi
    \disableopenany
  }
}

\title{Hardware Fingerprinting for the ARINC 429 Avionic Bus}

\author{
  \IEEEauthorblockN{
    Avishai Wool\IEEEauthorrefmark{1}, Nimrod Gilboa Markevich\IEEEauthorrefmark{2}
  }
  \IEEEauthorblockA{
    School of Electrical Engineering, Tel Aviv University\\
    Ramat Aviv, Israel\\
    Email: \IEEEauthorrefmark{1}yash@acm.org, \IEEEauthorrefmark{2}gmnimrod@gmail.com
  }
}

\begin{document}

\iftoggle{paper}{
  \maketitle
  \begin{abstract}
      ARINC 429 is the most common data bus in use today in civil avionics. 
However, the protocol lacks any form of source authentication. A technician with physical access to the bus is able to replace a transmitter by a rogue device, and the receivers will accept its malicious data as they have no method of verifying the authenticity of messages.

Updating the protocol would close off security loopholes in new aircraft but would require thousands of airplanes to be modified. For the interim, until the protocol is replaced, we propose the first intrusion detection system that utilizes a hardware fingerprinting approach for sender identification for the ARINC 429 data bus. Our approach relies on the observation that changes in hardware, such as replacing a transmitter or a receiver with a rogue one, modify the electric signal of the transmission. 

Because we rely on the analog properties, and not on the digital content of the transmissions, we are able to detect a hardware switch as soon as it occurs, even if the data that is being transmitted is completely normal. Thus, we are able to preempt the attack before any damage is caused.

In this \iftoggle{paper} {paper} {work} we describe the design of our intrusion detection system and evaluate its performance against different adversary models. Our analysis includes both a theoretical Markov-chain model and an extensive empirical evaluation. For this purpose, we collected a data corpus of ARINC 429 data traces, which may be of independent interest since, to the best of our knowledge, no public corpus is available. We find that our intrusion detection system is quite realistic: e.g., it achieves near-zero false alarms per second, while detecting a rogue transmitter in under 50ms, and detecting a rogue receiver in under 3 seconds. In other words, technician attacks can be reliably detected during the pre-flight checks, well before the aircraft takes off.
  \end{abstract}
}{
  \pagenumbering{gobble}  
  \input{title_page_english_outer.tex}
  \input{title_page_english_inner.tex}
  \frontmatter  
  \setfancyheaderfooter
  
  \tableofcontents
  \listoffigures
  \listoftables
  \mainmatter  
}

\level{Introduction}
\sublevel{Background}
  ARINC 429 \cite{arinc2004arinc429} is a prominent standard for wired intra-vehicle communication in civil aviation. Most active and retired airplanes contain ARINC buses \cite{fuchs2012evolution}, interconnecting the many digital systems that are necessary for the operation of an aircraft: sensors, radars, engines, cockpit controls and more.
  
  Safety and reliability are key objectives in avionics \cite{fuchs2012evolution}. Therefore, the main requirements of airborne subsystems are high determinism and low response times \cite{thanthry2005aviation}. ARINC 429 was designed accordingly. Security on the other hand, as we understand it today, was not a primary concern. At the time of the protocol's release in 1977, awareness of the threat of cyber-physical attacks was not as widespread as it is today. ARINC 429 was designed without any security features, such as encryption or source authentication, that are perceived today as essential to secure communication. In the years that have passed the importance of proper cybersecurity was demonstrated in numerous fields, from industrial networks \cite{langner2011stuxnet} to cars \cite{miller2015remote} to avionics \cite{costin2012ghost}. A recent study \cite{smith2020view} has found that attacks on wireless safety-related avionics systems have the potential of disrupting ongoing flights, inducing financial loss to airlines and reducing safety. In contrast to advancements in cybersecurity, there were no major revisions of the ARINC 429 standard since 1980 \cite{18937420070101}.
  
  AFDX is the successor to ARINC 429. It is Ethernet based, so IPSec or other modern solutions can be applied. Still, to quote from \cite{fuchs2012evolution}: ``[...] 429 will most likely not simply vanish; it will still be used in scenarios where simple signaling is sufficient, and in latency critical scenarios. It is a proven and extremely reliable technology and thus is also used as fallback network for the AFDX network, e.g. in the Airbus A380.''
  
  In particular, ARINC 429 has no mechanism for source authentication, so once an adversary has gained physical access to bus, any data they transmit will be accepted.
   
  One way to add authentication without an industry-wide update of the protocol is to implement it at a higher layer of the protocol stack. Unfortunately, in ARINC 429 there are only 19 data bits in a message. This is typically insufficient for a secure implementation of message code authentication (MAC).
  
  Intrusion detection systems (IDS) are often employed to retrofit security into similar systems that were designed without security in mind, such as CAN bus in automobile systems \cite{muter2011entropy}. An IDS is a software or device that continuously compares the observed behavior of the system to the expected behavior and raises an alarm whenever anomalies are detected. However, intrusion detection systems that analyze only the message content are limited in their ability to identify the sender of a message.  An attacker that is aware of the presence of the IDS might attempt to fool it by sending data that is identical or very similar to normal data.  Therefore, in order to identify the message source, we need to analyze the electrical signal. The guiding principle is that every transmitter is unique, even those of the same model and maker, due to minor defects in production, component tolerances etc., which manifest in the electrical signal. Hence, every signal has inimitable characteristics that can be used to identify the sender. To a lesser extent, the signal is also affected by the receivers and the bus topology. The characteristics can be used to identify changes to passive components on the bus. The process of learning to associate a signal to a transmitter is called hardware fingerprinting.
  
  It is important to note that hardware fingerprinting cannot replace the use of content-based IDSs, as hardware fingerprinting only detects attacks, where one transmitter sends messages, that are normally sent by another transmitter. On the other hand, an IDS that inspects only the data is at risk of being fooled by a clever adversary that mimics normal behavior. Using both methods in tandem supplies protection against the widest range of attacks. Exploring content-based detection is beyond the scope of this \iftoggle{paper} {paper} {work}.
  

\sublevel{Related Work} \label{RelatedWork}
  To the best of our knowledge, this is the first academic research to suggest hardware fingerprinting in ARINC 429. However, hardware fingerprinting was explored previously in a number of different domains: Ethernet \cite{kohno2005remote, uluagac2013passive, gerdes2012physical}; wireless radio \cite{ellis2001characteristics, brik2008wireless, xu2015device}; smartphone accelerometers, gyroscopes, microphones and cameras \cite{dey2014accelprint, das2016tracking}.
  
  One domain in particular is of special interest to us: controller area network (CAN bus) \cite{bosch1991canbus}, the most commonly used standard for in-vehicle communication in the automotive industry. ARINC 429 and CAN bus have a lot in common: Both are protocols for wired local area networks. Both are meant to be used in a static topology. Both share similar bit rates (up to 100 Kbits/sec in ARINC, up to 1 Mbits/sec in CAN) and similar word lengths (32 bits in ARINC, up to 128 bits in CAN). Both were formulated more than 30 years ago, and both were not designed for security but rather for safety, and as a consequence lack source authentication.
  
  In recent years a number of successful cyber-attacks were demonstrated on cars \cite{miller2015remote}, motivating researchers to search for new ways to hinder attacks.
  
  
  In \cite{cho2016fingerprinting} the authors propose using the measured arrival time of periodic messages to estimate characteristics of a transmitter's internal clock, which are then used as fingerprints. It has been demonstrated in \cite{sagong2018cloaking} that timing features can be emulated by an adversary, and therefore might not be a reliable means for identification.
  
  In \cite{murvay2014source} the authors demonstrate that transmitters of CAN messages can be identified by the signal's electrical characteristics. They propose using the CAN message ID field of the electric signal as a fingerprint. They also observed that the electrical characteristics remain stable for a period of months in a lab setup.
  
  In \cite{choi2018identifying} the authors suggest using time domain and frequency domain features extracted from the CAN ID field as fingerprints. They perform feature selection using the mutual information criterion in order to reduce the number of used features. Machine learning (ML) techniques are utilized for classification (SVM, NN and BDT).
  
  In \cite{cho2017viden} the authors devise a method for generating fingerprints based on the order statistics of voltage levels. The algorithm adapts to fluctuations in power supply levels and to changes in temperature by constantly updating the fingerprints based on new samples.
  
  In \cite{choi2018voltageids} the authors construct the fingerprints by extracting time domain and frequency domain features from selected parts of a message. They perform feature selection with sequential forward selection, and SVM and BDT for classification. Incremental learning techniques \cite{diehl2003svm} are employed in order to compensate for temporal changes of the characteristics.
  
  In \cite{kneib2018scission}, when identifying the source of a frame, the authors first construct a number of artificial signals from that frame. The artificial signals are made by cutting the signal and concatenating parts that share a similar behavior: positive slope transient, negative slope transient and stable positive voltage. For each of these three artificial signals, a set of time domain and frequency domain features are extracted.
  
  CAN bus and ARINC 429 use different line protocols, therefore methods presented in the above papers cannot be directly applied to our problem without change. A key difference between the two protocols is the bus topology. In CAN bus dozens of transceivers may share one bus. The main threat CAN papers are dealing with is device hijacking, where one ECU is remotely hacked, and starts to transmit the messages of another ECU on the same bus. Since during normal operation of a car, devices do not spontaneously join or leave the network, all the devices are known to the defender in advance. This scenario naturally fits into a multiclass classification setting, where a message needs to be identified as belonging to one of many known classes. In ARINC 429 on the other hand, only one transmitter is allowed on a line. Only that single transmitter is known to the defender beforehand. The task is to categorize a message as either normal or as an anomaly (an attack). Multiclass classification algorithms are not suitable for this task, because only samples from the normal class can be obtained for training. Instead we need to use anomaly detection algorithms \cite{pimentel2014review}. Of the CAN bus papers we reviewed, only \cite{choi2018identifying} and \cite{choi2018voltageids} extend their algorithms to handle detection of unknown transmitters.
  
\sublevel{Contributions}
  We propose the use of hardware fingerprinting in order to imbue ARINC 429 buses with source authentication capabilities. Applying the method only requires the attachment of a standard-compliant monitoring unit to the bus. This method does not require hardware or software updates to existing systems and is compliant with the current version of the ARINC standard.
  
  We describe the adversary models that our method is effective at protecting against. We then design an intrusion detection system with hardware fingerprinting capabilities, and evaluate its performance in these attack scenarios.
  
  We explore the ability to distinguish between devices from different vendors and between devices of the same model, based on the hardware fingerprints of individual transmitted words. We find that it is possible to distinguish between transmitters and receivers by their electric signal, with low error rates. This observation applies both to devices from different vendors and to devices from the same vendor and model, which are supposedly identical.
  
  We explore the effect of receivers and transmission lines on performance. We see that adding a receiver does not yield a significant change in the signal. However, switching a receiver by another receiver, when combined with a change to the transmission line, is detectable by our method.
  
  We compare different feature sets under different the adversarial models. Somewhat surprisingly, we find that using the raw samples, without extracting any features, yields the best outcome when detecting a transmitter switch. In case of a receiver switch, we find that features derived from a polynomial fit outperform the other feature set.
  
  In order to drive the false-alarms-per-second rate to zero, we suggest to augment the per-word anomaly detection by a ``suspicion counter'' that increases with each word flagged as an anomaly, and decreases with every normal word. We first analyze the suspicion counter using a Markov-chain model, and then evaluate the full system's performance using the empirical data. 
   We demonstrate that our intrusion detection system is quite realistic: e.g., it achieves near-zero false alarms per second, while detecting a rogue transmitter in under 50ms, and detecting a rogue receiver in under 3 seconds. In other words, technician attacks can be reliably detected during the pre-flight checks, well before the aircraft takes off.

  \textbf{Organization}: Section \ref{Preliminaries} describes the ARINC 429 protocol and the adversary model. 
  Section \ref{TheDataSet} presents the data set we collected,
  and Section \ref{PreliminaryTests} demonstrates a proof of concept.
  Section \ref{Approach} outlines our hardware fingerprinting approach. Section \ref{SignalSegmentation} describes our signal segmentation process and Section \ref{FeatureSets} covers the various feature sets we considered. 
  Section \ref{PerformanceEvaluationSingleWord} describes the empirical evaluation of our detection method when it is based on single words. Section \ref{ModelingSuspicionCounter} presents a Markov-chain analysis of the suspicion counter, and Section \ref{PerformanceEvaluationCompleteMethod} describes the empirical performance evaluation of the complete method. We conclude with Section \ref{Conclusions}.
  
\level{Preliminaries} \label{Preliminaries}
\sublevel{The ARINC 429 Standard}
  ARINC Specification 429 \cite{arinc2004arinc429}, also named ``Mark 33 Digital Transfer System (DITS)'', is a standard of the avionics industry. It defines a protocol for the communication between avionics system elements over a local area network. First published in 1977 by Aeronautical Radio, Inc., it has since become one of the most widely used data bus protocols in civil aircrafts \cite{MoirIan2013DBN}. The protocol encompasses different layers: from the physical requirements, through the electronic characteristics of the signal, data format and ending with a file transfer technique.

\begin{figure}[t]
    \centering
    \includegraphics[width=1.0\linewidth, angle=0]{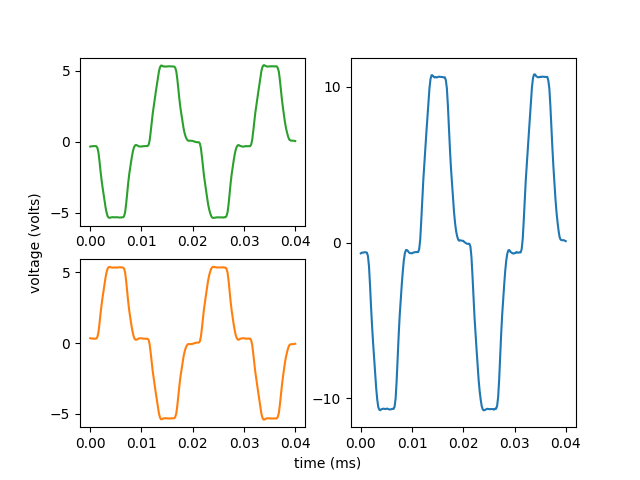}
    \caption{ARINC 429 bus showing the voltage difference between twisted pair for bits 0101. Counter-clockwise starting from the top left: Line A, Line B, the differential signal}
    \label{fig:word_example}
  \end{figure}

  We continue with a short description of those parts of the specifications, which are relevant to this \iftoggle{paper} {paper} {work}. In ARINC 429 the communicating entities are called line-replaceable units (LRU). Data is transmitted over a single twisted and shielded pair of wires. The cable shield is grounded on both ends. The lines are denoted Line A and Line B. Differential signaling is used, meaning that the signal is the voltage difference from Line A to Line B, rather than the difference from one wire to ground. Bipolar return-to-zero (BRTZ) modulation is used as a line protocol. BRTZ is a tri-level state modulation: we refer to the three voltage levels as ``HI'', ``LO'' and ``NULL''. A binary 1 is encoded as a positive voltage pulse ``HI'', and a binary 0 is encoded as a negative voltage pulse ``LO''. In between transmissions, the voltage drops to 0V, ``NULL''. Every ``HI'' and every ``LO'' are preceded and are followed by a ``NULL'', even if repeating bit values are transmitted consecutively. The differential output voltage from line A to line B is $10V \pm 1$ in ``HI'' mode, $0 \pm 0.5$ in ``NULL'' mode and $-10V \pm 1$ in ``LO'' mode.  Figure \ref{fig:word_example} shows a recording of a transmission on an ARINC 429 data bus.
  
  Data is transmitted in words that are 32-bit long. The bits are transmitted in the following order, from first to last: 8, 7, ..., 2, 1, 9, 10, ..., 32. This order is a legacy from older systems. In this \iftoggle{paper} {paper} {work}, words are interpreted as though an MSB-first transmission order is in place.
  
  Data on the ARINC 429 bus is transmitted unidirectionally from a single transmitter LRU to up to 20 receiver LRUs. Only one transmitter LRU is allowed on the bus - a separate bus is required for each transmitter. Since there is only one transmitter on each bus, there is no sender ID field in ARINC messages.
   
  The protocol allows a choice of one of two bit rates: Slow, at 12.0 to 14.5 Kbits/sec, and fast, at 100 Kbits/sec. The bit rate on a bus is fixed and maintained within \%1. The signal is self-clocking.
  
  MIL-STD-1553 \cite{united1986milstd1553} is the military bus standard alternative of ARINC 429.
  
\sublevel{The Adversary Model}
  Our method is designed to guard against ``technician attacks''. This type of attack involves an adversary that has brief physical access to the system. Such an adversary is able to replace LRUs or add new ones to the bus.
  
  The adversary may have prior knowledge of the hardware and topology of the attacked system. The reverse is not true: As defenders, we have no prior knowledge of what the adversary's hardware might be. However, we do assume that the adversary will use commercial off-the-shelf hardware.
  
  We only consider attacks, where the adversary changes the hardware that is connected to the bus, as other types of attack do not affect the signal characteristics.
  
  We distinguish between several types of attacks.

\subsublevel{A Rogue Transmitter}
  In this type of attack an adversary replaces a legitimate transmitter LRU by a rogue one. During an initial dormant phase of the attack, the new device imitates the behavior of the original transmitter, transmitting data exactly as requested, in order to remain hidden. Only at a later time, the attack moves on to its active phase. During this phase the rogue transmitter LRU sends out messages which are meant to disrupt the work of the system, and in extreme cases causes irreversible damage to the electronic or physical components.
  

\subsublevel{A Rogue Receiver}
  In this attack type of attack the adversary replaces a legitimate receiver LRU by a rogue one, or adds a rogue receiver LRU without detaching another LRU. By doing this the adversary gains access to the transmitted data, which might be otherwise inaccessible, and may use this data to cause harm through another attack channel.

\subsublevel{Adding a Transmitter or Converting a Receiver to a Transmitter}
  An attack wherein the adversary adds another transmitter LRU to the bus, without detaching the legitimate transmitter, is actually not possible to perform on the ARINC bus. The ARINC 429 bus is designed to allow exactly one transmitter LRU. Connecting two transmitters to the same bus irreparably violates the electrical properties of the system. Therefore, an adversary cannot simply add a transmitter (built from off-the-shelf components) to the bus. 
  The adversary needs to make sure that the legitimate transmitter is disconnected before connecting the rogue transmitter. Otherwise, there is a risk that data will fail to be delivered. In fact, when we naively connected two transmitters to the same bus, the peak to peak voltage dropped by half, and the legitimate communication on the bus failed. While we don't assert that this will always be the case, it serves as a cautionary anecdote for adversaries. Further, an adversary may possibly construct special hardware that would allow the bus to function with two or more transmitters, for example by disconnecting the legitimate transmitter during transmissions of the rogue transmitter, but the fact remains that standard commercial components would not suffice.

  Furthermore, it is not possible to turn a receiver LRU into a transmitter LRU by hijacking its software, since the LRU's wiring does not permit it. 
 
\level{The Data Set} \label{TheDataSet}
  To the best of our knowledge, there is no publicly available data set that contains high rate voltage samples of ARINC 429 protocol. We gathered our own data set, with the kind assistance of \textit{Astronautics C.A. LTD.} \cite{astronautics2019home}.
  
  We sampled two types of transmitters:
  \begin{enumerate}
     \item An M4K429RTx test equipment from \textit{Excalibur Systems} \cite{excalibur2019m4k429rtx}. The Excalibur equipment hosts two transmitters which we label \(\text{E}_1\) and \(\text{E}_2\).
     \item ADK-3220 Evaluation boards, manufactured by \textit{Holt Integrated Circuits INC.} \cite{holt2019evaluation}. The board contains a HI-3220PQ ARINC 429 Protocol IC connected to 8 HI-8597PSIFF line drivers chips. We use 4 of the transmitters and two different boards. We label the transmitters \(\text{H}_{xy}\), where x is the board number, 1 or 2, and y is the transmitter number from 0 to 3.
  \end{enumerate}
  
  The transmitters were connected to one or more of the following receivers:
  \begin{enumerate}
    \item An EDCU, a proprietary device manufactured by \textit{Astronautics C.A. LTD.} \cite{astronautics2019edcu}. The device has two receivers which we label \(\text{P}_1\) and \(\text{P}_2\).
    \item The ADK-3220 Evaluation boards also host 16 integrated line receivers. We use 2 of the ports with the two boards. We label the receivers the same way as the transmitters with \(\text{H}_{xy}\), where x is the board number as before, and y is the receiver number.
  \end{enumerate}
  
  For sampling we used a Keysight DSO9254A scope. All signals were sampled at 50Msa/s at a scope bandwidth of 25MHz. The probes are 500MHz, 10M\(\Omega\), 11pF. Each line was sampled individually. We further downsampled digitally by a factor of 10 to a rate of 5 MSa/s using a 30 point FIR filter with Hamming window.
  
\begin{figure}[t]
    \centering
    \includegraphics[width=1.0\linewidth, angle=0]{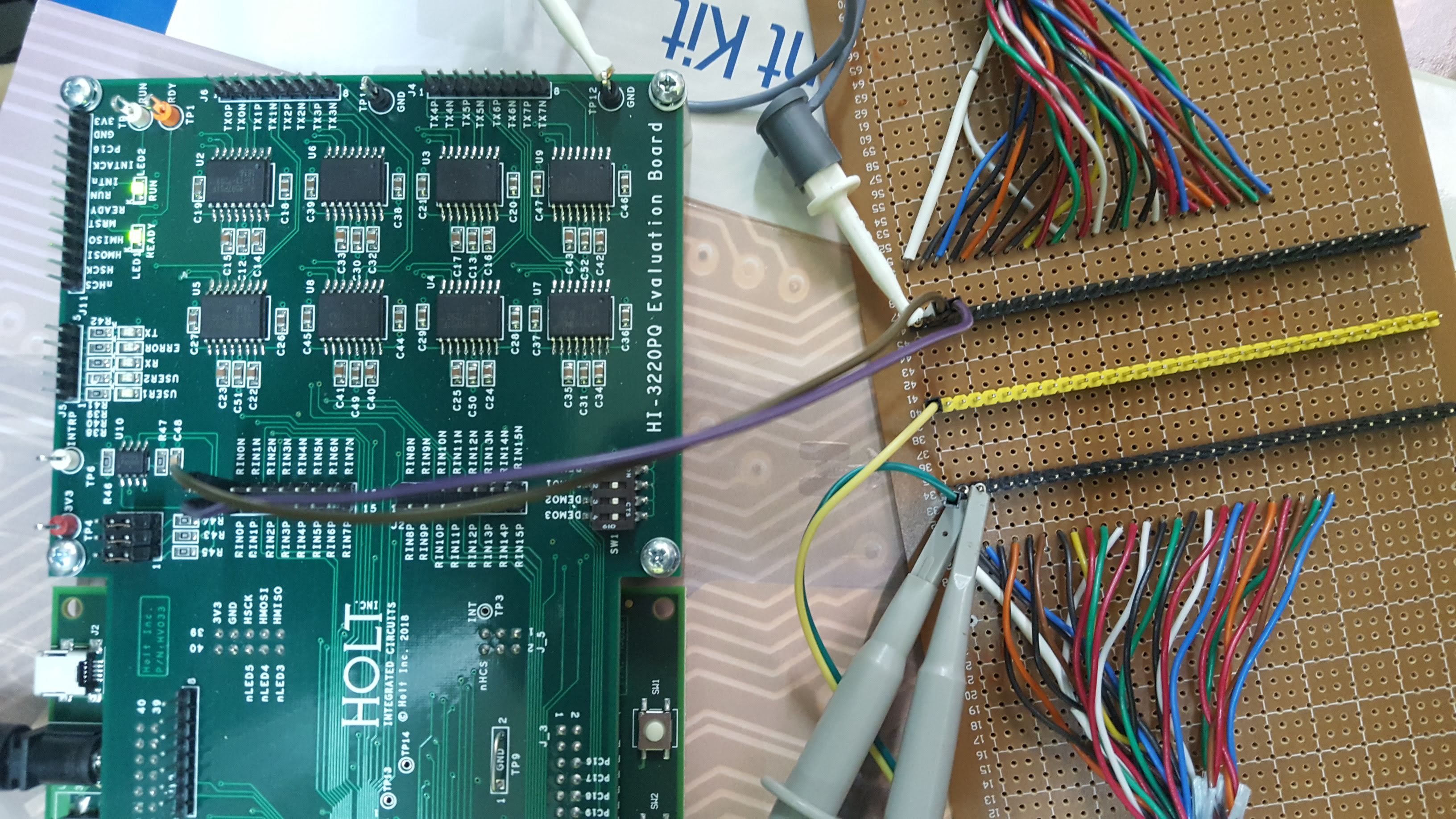}
    \caption{The Holt evaluation board on the left, and the fabricated connector board on the right}
    \label{fig:SetupImage}
  \end{figure}

  The transmitters and receivers were connected through a custom board that exposes the wires, which we fabricated for this purpose (see Figure \ref{fig:SetupImage}).
  
  All the devices transmitted the same data at a bit rate of 100 Kbits/sec. 6 values of words were transmitted. Interpreting the words with MSB-first transmission order, the values are: \texttt{0x00000000}, \texttt{0xFFFFFFFF}, \texttt{0x55555555}, \texttt{0xAAAAAAAA}, \texttt{0x5A5A5A5A}, \texttt{0xA5A5A5A5}. Note that these words include all the possible segment types. By transmitting the same data on all devices we make sure that in our experiments the IDS cannot unintentionally use the message content to make its decisions.
  
  In addition to the recordings from different transmitter-receiver pairs, we recorded \(\text{E}_1\) and \(\text{E}_2\) transmitting to \(\text{P}_1\) and \(\text{P}_2\) respectfully, with different Holt devices attached as additional receivers.
  
  Table \ref{tab:RecordingsSummery} shows the different combinations of transmitter-receiver in our data set, and the number of words recorded for each combination.
  
  \begin{table}
    \caption{Distribution of Recorded Words in the Data Set}
    \label{tab:RecordingsSummery}
    \centering
    \begin{tabular}{|c | c c c|} 
      \hline
      Row \# & Transmitter & Receiver & \#Words \\ [0.5ex] 
      \hline\hline
      1 & \(\text{E}_1\) & \(\text{P}_1\) & 4920 \\ 
      \hline
      2 & \(\text{E}_1\) & \(\text{P}_1\) \& \(\text{H}_{10}\) & 4920 \\
      \hline
      3 & \(\text{E}_1\) & \(\text{P}_1\) \& \(\text{H}_{12}\) & 4920 \\
      \hline
      4 & \(\text{E}_1\) & \(\text{P}_1\) \& \(\text{H}_{20}\) & 4920 \\
      \hline
      5 & \(\text{E}_1\) & \(\text{P}_1\) \& \(\text{H}_{22}\) & 4920 \\
      \hline
      6 & \(\text{H}_{10}\) & \(\text{P}_1\) & 4920 \\
      \hline
      7 & \(\text{H}_{11}\) & \(\text{P}_1\) & 4920 \\
      \hline
      8 & \(\text{H}_{12}\) & \(\text{P}_1\) & 4920 \\
      \hline
      9 & \(\text{H}_{13}\) & \(\text{P}_1\) & 4920 \\
      \hline
      10 & \(\text{H}_{20}\) & \(\text{P}_1\) & 4920 \\
      \hline
      11 & \(\text{H}_{21}\) & \(\text{P}_1\) & 4920 \\
      \hline
      12 & \(\text{H}_{22}\) & \(\text{P}_1\) & 4920 \\
      \hline
      13 & \(\text{H}_{23}\) & \(\text{P}_1\) & 4920 \\
      \hline
      14 & \(\text{E}_2\) & \(\text{P}_2\) & 4920 \\ 
      \hline
      15 & \(\text{E}_2\) & \(\text{P}_2\) \& \(\text{H}_{10}\) & 4920 \\
      \hline
      16 & \(\text{E}_2\) & \(\text{P}_2\) \& \(\text{H}_{12}\) & 4920 \\
      \hline
      17 & \(\text{E}_2\) & \(\text{P}_2\) \& \(\text{H}_{20}\) & 4920 \\
      \hline
      18 & \(\text{E}_2\) & \(\text{P}_2\) \& \(\text{H}_{22}\) & 4920 \\
      \hline
      19 & \(\text{H}_{10}\) & \(\text{P}_2\) & 4920 \\
      \hline
      20 & \(\text{H}_{11}\) & \(\text{P}_2\) & 4920 \\
      \hline
      21 & \(\text{H}_{12}\) & \(\text{P}_2\) & 4920 \\
      \hline
      22 & \(\text{H}_{13}\) & \(\text{P}_2\) & 4920 \\
      \hline
      23 & \(\text{H}_{20}\) & \(\text{P}_2\) & 4920 \\
      \hline
      24 & \(\text{H}_{21}\) & \(\text{P}_2\) & 4920 \\
      \hline
      25 & \(\text{H}_{22}\) & \(\text{P}_2\) & 4920 \\
      \hline
      26 & \(\text{H}_{23}\) & \(\text{P}_2\) & 4920 \\
      \hline
    \end{tabular}
  \end{table}
 
 \level{Preliminary Tests} \label{PreliminaryTests}
 
  \begin{figure}[t]
    \centering
    \includegraphics[width=1.0\linewidth, height=0.4\textheight,angle=0,keepaspectratio]{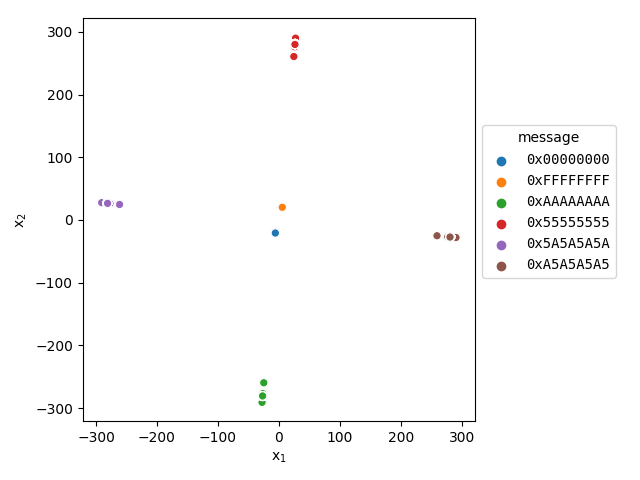}
    \caption{PCA dimensionality reduction of whole words}
    \label{fig:pca_by_message}
  
    \centering
    \includegraphics[width=1.0\linewidth, height=0.4\textheight,angle=0,keepaspectratio]{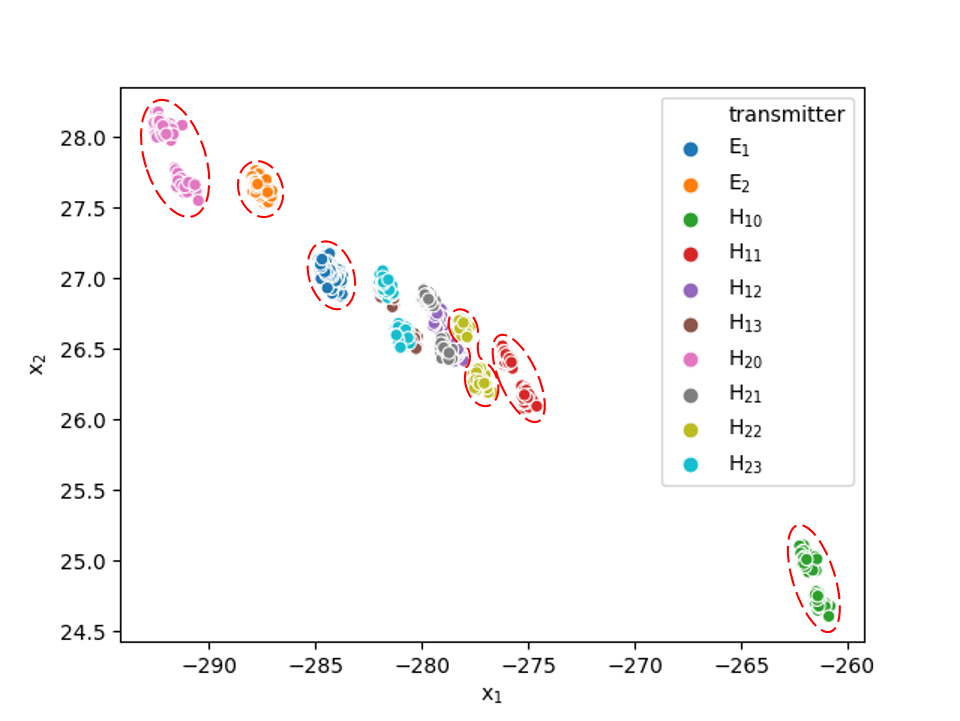}
    \caption{Dimensionality reduction of whole words, message \texttt{0x5A5A5A5A}, label by transmitter ID}
    \label{fig:pca_by_tx_id}
  \end{figure}
  
  \begin{figure}[t]
    \centering
    \includegraphics[width=1.0\linewidth, height=0.4\textheight,angle=0,keepaspectratio]{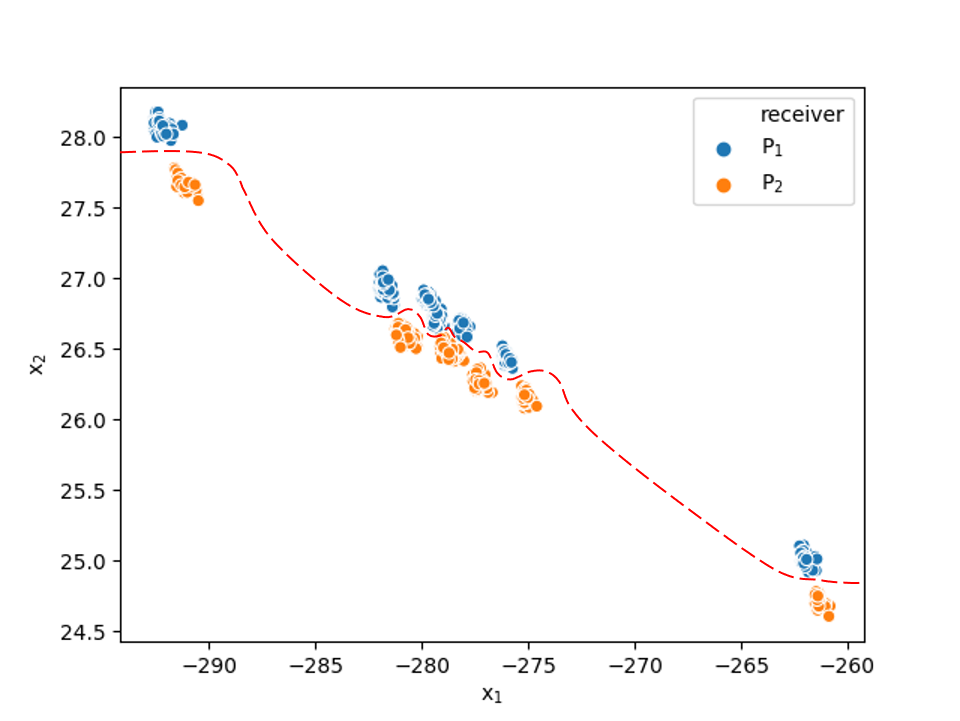}
    \caption{Dimensionality reduction of whole words, message \texttt{0x5A5A5A5A}, label by receiver ID}
    \label{fig:pca_by_rx_id}
    
    \centering
    \includegraphics[width=1.0\linewidth, height=0.4\textheight,angle=0,keepaspectratio]{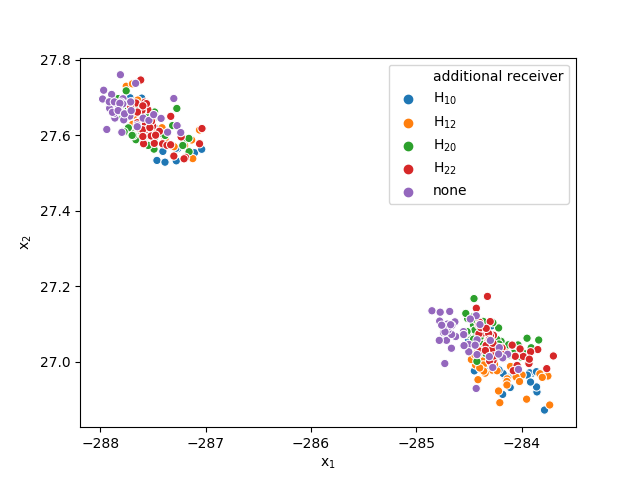}
    \caption{Dimensionality reduction of whole words, message \texttt{0x5A5A5A5A}, label by additional receiver ID}
    \label{fig:pca_by_load}
  \end{figure}

  Before we started evaluating our IDS, we performed a basic proof of concept investigation to assess the ``detectability'' of hardware changes. To do so we reduced raw traces of whole words (about 1600 samples) to 2 dimensions with Primary Component Analysis (PCA) \cite{pearson1901liii}.
 
  Figure \ref{fig:pca_by_message} shows the result of the PCA on a representative subset of our data set. In the figure there are 6 distinctive clusters, each one corresponds to words with a different message. The PCA naturally separates the words by the value of the message they transmit, since this property is responsible for the bulk of the difference between words.
  
  In Figure \ref{fig:pca_by_tx_id} we zoom in on the leftmost cluster, which contains only words with the message \texttt{0x5A5A5A5A}. We see that in fact, the \texttt{0x5A5A5A5A} cluster is made up of smaller clusters. The words are labeled according to the transmitter that sent them. The figure shows that most (but not all) groups of words from the same transmitter form a cluster that is visibly distinguishable from the others, as demonstrated by the dashed ellipses. This indicates that the information needed to distinguish between different transmitters is present in the sampled words.
  
  Figure \ref{fig:pca_by_rx_id} shows the same words, labeled according to the primary receiver that was connected to the line ($P_1$ or $P_2$). The transmitters $E_1$ and $E_2$ are removed from the figure because each was only recorded with one receiver. We see that the two labels are visually distinguishable, as demonstrated by the dashed curved line. We conclude that receivers, as well as transmitters, can be fingerprinted.
  
  Lastly, in Figure \ref{fig:pca_by_load} the words are labeled according to the presence and identity of an \textit{additional receiver} on the bus. Only words from transmitters $E_1$ and $E_2$ are displayed, because the Holt transmitters were not recorded with additional receivers (recall Table \ref{tab:RecordingsSummery}). The figure shows that identifying an additional receiver is much more difficult than identifying a transmitter or a receiver, since we cannot separate words from different labels by a simple line or convex shape. This fact does not mean that it is impossible to detect an additional transmitter, only that it requires more information than that which is present in the first two primary components alone.
  
  Of course, we can't use this basic PCA method to implement an anomaly detector. First, we would need to record in advance all the possible words the transmitter LRU might transmit. And second, we used samples from all recordings when calculating the primary components, without distinguishing between a normal and a rogue system---in a real IDS, we can only use samples from the guarded bus to train our anomaly detector. Still, the above figures provide positive evidence that transmitter switches and receiver switches change the electrical signal in a noticeable way, while they do not provide definite proof that receiver addition does not.

\level{The Hardware Fingerprinting Approach} \label{Approach}
  The fingerprinting IDS we propose has to be attached to the bus it is guarding. During a training period it samples the bus and learns the transmitter LRU's characteristics. We assume that during this time only legitimate devices are present on the bus. We further assume that access to the IDS is restricted, so that only authorized personnel are able to trigger the training mechanism. This restriction is in place in order to prevent an adversary from retraining the IDS after switching the guarded transmitter by the rogue one.
  
\sublevel{Advantage Over Content Based IDSs}
  An intrusion detection system which draws its information exclusively from the digital content of the transmitted messages will be unable to detect the rogue transmitter LRU during the latter's dormant phases, or to detect a rogue receiver LRU. Only during an active phase of the attack of a rogue transmitter LRU is the transmitted data distinguishable from that of a legitimate transmission. By utilizing information from the analog domain, we are able to detect hardware changes independently from the transmitted data.  Usually, before takeoff, the aircraft systems are checked for basic integrity. During these pre-flight operations the changes to the bus can be detected, even if the transmitter LRU is sending normal data.
  
\sublevel{IDS Overview} \label{Overview}
  We will next describe our proposed method of anomaly detection. We divide the algorithm into several stages. This section provides an overview of these steps. In the subsequent sections selected stages are explained in greater detail as needed.
  
  \begin{enumerate}
    \item \textbf{[Acquisition]}
          We sample both lines of the bus at a sampling rate that is 50 times higher than the bit rate. We used a sample rate of 5 MSa/s. The differential signal is obtained by subtracting the samples of line B from the samples of line A.
    \item \textbf{[Segmentation]}
          Each word is split into 127 segments of 10 different types, based on voltage levels. The purpose of the segmentation is to eliminate the effect of the transmitted data, i.e., the content of the word, on the final decision of the anomaly detector. See Section \ref{SignalSegmentation} for details.
    \item \textbf{[Feature Extraction]}
          We extract multiple features from each segment. See Section \ref{FeatureSets} for details.
    \item \textbf{[Anomaly Detection per Segment]}
          The features from each segment are fed into a trained anomaly detector. Each \textit{segment} is marked as either ``normal'' or ``anomaly''.
    \item \textbf{[Voting]}
          A word is flagged as an ``anomaly'', if the number of ``normal'' segments it contains does not exceed $T_{votes}$, a calibrated threshold.
    \item \textbf{[Suspicion Counter]}
          We keep a counter of anomalous words. Each time a word is declared as an ``anomaly'', the counter is increased by 1, and each time a word is declared as ``normal'', the counter is decreased by 1, to a minimum of 0. Once the counter reaches a threshold of $T_{suspicion}$ an alarm is raised.
  \end{enumerate}
  
\sublevel{Anomaly Detection per Segment}
  Our basic building block uses per-segment anomaly detection. As we shall see there are 10 types of segments, as detailed in Table \ref{tab:SegmentationLevels} in Section \ref{SignalSegmentation}. A segment's characteristics depend on its type. Therefore, we opted to train a different anomaly detector for each type of segment.
  
  Anomaly detection, sometimes called novelty detection, is a well-established. There are numerous outlier and anomaly detection algorithms available in the literature such as K-Nearest Neighbors \cite{hautamaki2004outlier}, Mixture Models \cite{paalanen2006feature}, One-Class SVM \cite{scholkopf2000support} and Isolation Forest \cite{liu2008isolation}. An extensive review of various algorithms is presented in \cite{pimentel2014review}.
  
  For the anomaly detection task, we chose to work with the Local Outlier Factor (LOF) by Breunig et al.\ \cite{breunig2000lof}. LOF was shown to work better than other algorithms for the task of network intrusion detection\cite{lazarevic2003comparative}. This fact, together with the available scikit-learn \cite{scikit-learn} Python implementation, made it an appealing choice. Comparing different anomaly detection algorithms is beyond the scope of this \iftoggle{paper} {paper} {work}.
  
  LOF is a density-based outlier detection algorithm. According to the LOF algorithm, an outlier is defined as a data point (feature vector), whose local density is greater than the average of local densities of its neighbors by a large enough margin. A local density of a data point is the inverse of the average distance of the point from its neighbors.
  
  There are several hyper-parameters for the LOF algorithm. In all cases we used the default parameters provided by the implementation. For the number of neighbors examined when calculating the LOF the default is 20. We used the Euclidean metric for the distance measure. The threshold on the local outlier factor that defines an anomaly is automatically set so that 10\% of samples in the \textit{training set} are outliers.
  
  We constructed a separate anomaly detector for each type of segment. Each segment is fed individually into its appropriate LOF anomaly detector. The LOF outputs its determination regarding the source of the segment, either ``normal'' or ``anomaly''.
  
\sublevel{Voting}
  We gather the decisions made by the different LOF detectors for all segments of the same word. The number of segments that have been identified as normal is subjected to a voting threshold $T_{votes}$. If it does not exceed $T_{votes}$, the word is flagged as an ``anomaly'', otherwise, it is flagged as ``normal''.
  
\sublevel{Suspicion Counter}
  According to our adversary model, an attacker tampers with the system only once. Therefore, we expect the true label of all words in the incoming stream to be identical---either all the words originate from the original system, or all the words originate from a compromised system. We utilize this attack model to reduce the probability of making an error. Taking note from \cite{kneib2018scission} we incorporate an anomaly counter, which we name the suspicion counter.
  
  The suspicion counter is a counter that is updated on the arrival of a new word. The initial value of the counter is 0. When a word is declared as an ``anomaly'', the counter is increased by 1, and when a word is declared as ``normal'', the counter is decreased by 1, to a minimum of 0. Once the counter reaches a calibrated threshold of $T_{suspicion}$ an alarm is raised.
  
\level{Signal Segmentation} \label{SignalSegmentation}
  Our method aims to rely solely on the physical characteristics of the hardware, and aims to be completely agnostic to the transmitted data. In order to achieve this goal, we divide each word into sub-bit non-overlapping segments.
  
  In a BRTZ line protocol, each bit comprises of 4 distinct segments. For example, a `1' bit starts with a transition up from ``NULL'' to ``HI'', then a plateau on ``HI'', then a transition down from ``HI'' back to ``NULL'', and finally a ``NULL'' plateau. Furthermore, we observed 4 different variants of ``NULL'', depending on the current and on the next bit. E.g., a ``NULL'' between two `1' bits tends to be ``smile''-shaped, while a ``NULL'' between two `0' bits has a ``frown'' shape. All in all, we identified 10 different segment types, as listed in Table \ref{tab:SegmentationLevels}.
  
  Thus, we split every 32-bit word into 127 segments. Note that there are only 127 segments, not 128, because the last bit is followed by a long ``NULL'' that lasts until the next word and has a unique shape. We do not associate this segment with any word.
  
  The segmentation is performed in the following manner. A segment starts where the voltage level of the signal rises above / falls below a certain threshold, and ends where it falls below / rises above another threshold. 4 different thresholds are employed in order to produce a stabling hysteresis effect. We denote them as follows, and use them and their negative to define segment boundaries:
  
  \begin{align*}
    V_{l_1} = 2.0V \\
    V_{l_2} = 2.8V \\
    V_{h_1} = 8.0V \\
    V_{h_2} = 7.2V 
  \end{align*}
  
\begin{figure}[t]
    \centering
    \includegraphics[width=1.0\linewidth, angle=0]{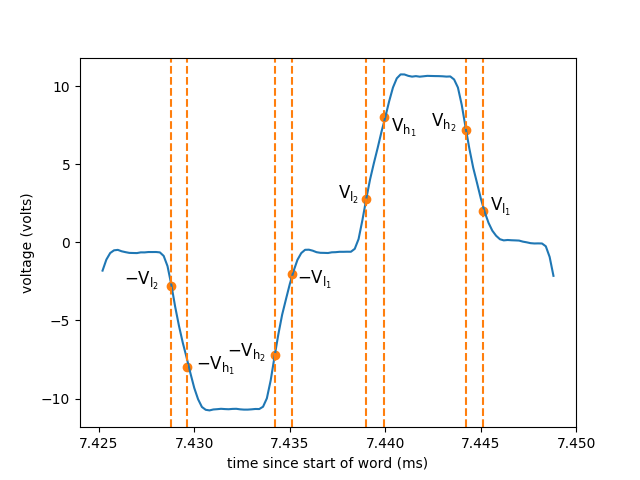}
    \caption{A segmentation example of the bits 01. The trace exhibits all 4 up/down transitions, the ``HI'' and ``LO'' plateaus, and 3 of the 4 possible ``NULL'' segment types}
    \label{fig:SegmentationTrace}
  \end{figure}

  Table \ref{tab:SegmentationLevels} shows the voltage levels used for each segment type. Figure \ref{fig:SegmentationTrace} shows an example of word segmentation.
  
  \begin{table}
    \caption{Voltage Thresholds per Segment Type}
    \label{tab:SegmentationLevels}
    \centering
    \begin{tabular}{|c c c|} 
      \hline
      Segment & Starting Threshold & Ending Threshold \\ [0.5ex] 
      \hline\hline
      LO & falls below $-V_{h_1}$ & rises above $-V_{h_2}$ \\
      \hline
      HI & rises above $V_{h_1}$ & falls below $V_{h_2}$ \\
      \hline
      NULL, HI to HI & falls below $V_{l_1}$ & rises above $V_{l_2}$ \\
      \hline
      NULL, HI to LO & falls below $V_{l_1}$ & falls below $-V_{l_2}$ \\
      \hline
      NULL, LO to LO & rises above $-V_{l_1}$ & falls below $-V_{l_2}$ \\
      \hline
      NULL, LO to HI & rises above $-V_{l_1}$ & rises above $V_{l_2}$ \\
      \hline
      Up from LO & rises above $-V_{h_2}$ & rises above $-V_{l_1}$ \\
      \hline
      Up from NULL & rises above $V_{l_2}$ & rises above $V_{h_1}$ \\
      \hline
      Down from HI & falls below $V_{h_2}$ & falls below $V_{l_1}$ \\
      \hline
      Down from NULL & falls below $-V_{l_2}$ & falls below $-V_{h_1}$ \\
      \hline
    \end{tabular}
  \end{table}
  
\level{Feature Sets} \label{FeatureSets}
  In our work we compare the performance of the feature sets described below.
 
{\bf Raw Time-Domain Samples.}
  This feature set consists of the raw vector of sequential voltage samples. The only additional operation we perform after segmentation is truncating the segments to a common length, since the LOF algorithm expects all data points to be vectors of the same dimension. The length varies depending on the segment type, as shown in Table \ref{tab:feature_set_sizes}. At the sample rate we use (recall Section \ref{Overview}) the number of samples per segment is quite low - between 4-24 (see Table \ref{tab:feature_set_sizes}). This makes the Raw feature set a practical choice.
  
{\bf Generic Time-Domain Feature Set.}
  As discussed in Section \ref{RelatedWork}, in recent years a number of papers suggested using extracted features to perform hardware fingerprinting \cite{dey2014accelprint, choi2018identifying, choi2018voltageids, kneib2018scission}. They all utilized time-domain features such as mean, standard deviation, skewness etc., with good results.
  
  These features are of a generic nature, in the sense that the shape of the signal does not affect the extraction process. Due to this property, it is possible to apply the same features in different fields of research.
  
  We use the features that were presented in \cite{kneib2018scission} as our Generic set. Six of the eight features in this feature set are used in all four cited papers. The features we used are listed in Table \ref{tab:generic_feature_set}.
  
  \begin{table}
    \caption{Generic Feature Set}
    \label{tab:generic_feature_set}
    \centering
    \begin{tabular}{|c c|} 
      \hline
      Feature & Description \\ [0.5ex] 
      \hline\hline
      Mean & \(\mu = \frac{1}{N}\sum_{i=1}^{N}x(i)\) \\
      \hline
      Standard Deviation & \(\sigma = \sqrt{\frac{1}{N}\sum_{i=1}^{N}(x(i)-\mu)^2}\) \\
      \hline
      Variance & \(\sigma^2 = \frac{1}{N}\sum_{i=1}^{N}(x(i)-\mu)^2\) \\
      \hline
      Skewness & \(skew = \frac{1}{N} \sum_{i=1}^{N}(\frac{x(i)-\mu}{\sigma})^3\) \\
      \hline
      Kurtosis & \(kurt = \frac{1}{N} \sum_{i=1}^{N}(\frac{x(i)-\mu}{\sigma})^4\) \\
      \hline
      Root Mean Square & \(rms = \sqrt{\frac{1}{N}\sum_{i=1}^{N}x(i)^2}\) \\
      \hline
      Maximum & \(max = max(x(i))_{i=1...N}\) \\
      \hline
      Energy & \(en = \frac{1}{N}\sum_{i=1}^{N}x(i)^2\) \\
      \hline
    \end{tabular}
  \end{table}
  
  In addition to time-domain features, the cited papers also employ frequency-domain features. We do not use frequency-domain features in this \iftoggle{paper} {paper} {work}. We argue that the non-periodic nature of the signals, that are the result of our segmentation method, does not benefit from frequency analysis.

{\bf Polynomial Feature Set.}
  The features in this set are calculated by performing a least squares polynomial fit and taking each coefficient as a separate feature, plus the residual as an additional feature.
  
  In order to avoid overfitting, we fit each type of segment with a polynomial function of an appropriate degree. For the four transitions (``Up from LO'', ``Up from NULL'', ``Down from HI'', ``Down from NULL'') we use a degree of 2. For ``NULL, HI to HI'' and ``NULL, LO to LO'' we use a degree of 6, on account of these segments being even functions. For the remaining segments we use a degree of 7 for similar reasons. Note that the number of features is always one more than the degree due to the residual.
  
{\bf Hand-Crafted Feature Set.}
  In this feature set there are different features for each segment type.
  
  We observed that the ``HI'' segments contain an overshoot followed by ripples. We denote by \((t_1, v_1), (t_2, v_2), (t_3, v_3)\) the time and voltage level at the first local maxima (the overshoot), then the first local minima that follows and then the first local maxima that follows. Time is measured from the beginning of the segment. The features we take are the above 6 values, in addition to the differences in time and voltage of the second and third points from the first point: \(t_2-t_1, v_2-v_1, t_3-t_1, v_3-v_1\). The features in the ``LO'' segments are a mirror image of the features in the ``HI'' segment.
  
  For ``NULL, HI to HI'' and ``NULL, LO to LO'' we only take the time and voltage levels at the overshoot \((t_1, v_1)\): not all segments of these types in the data set have ripples.
  
  The 4 transition segments are linear-like. For them we extract 2 features. The first is the mean of the first derivative. This quantifies the slope. The second is feature is the mean of differences of the segment from a line that passes between the segment's endpoints. This feature quantifies the deviation of the segment from a straight line.
  
  The segments ``NULL, LO to HI'' and ``NULL, HI to LO'' do not participate: not all segments of these types in the data set contain an overshoot.
  
  \begin{table}
    \caption{Number of Features per Segment Type}
    \label{tab:feature_set_sizes}
    \centering
    \resizebox{\columnwidth}{!}{
    \begin{tabular}{|c c c c c c|}
      \hline
      Segment & Segment Length & Raw & Generic & Polynomial & Hand-Crafted \\ [0.5ex]
      \hline\hline
      LO & 20-24 & 20 & 8 & 7 & 10 \\
      \hline
      HI & 20-23 & 20 & 8 & 7 & 10 \\
      \hline
      NULL, HI to HI & 17-22 & 17 & 8 & 7 & 2 \\
      \hline
      NULL, HI to LO & 17-21 & 17 & 8 & 8 & 0 \\
      \hline
      NULL, LO to LO & 17-22 & 17 & 8 & 7 & 2 \\
      \hline
      NULL, LO to HI & 17-21 & 17 & 8 & 8 & 0 \\
      \hline
      Up from LO & 4-6 & 4 & 8 & 3 & 2 \\
      \hline
      Up from NULL & 4-6 & 4 & 8 & 3 & 2 \\
      \hline
      Down from HI & 4-5 & 4 & 8 & 3 & 2 \\
      \hline
      Down from NULL & 4-5 & 4 & 8 & 3 & 2 \\
      \hline
    \end{tabular}
    }
  \end{table}

\level{Detection based on a Single Word} \label{PerformanceEvaluationSingleWord}
\sublevel{Methodology} \label{Methodology}
  In order to evaluate the performance of our algorithm, we performed an extensive series of experiments. In each experiment we selected one transmitter LRU as a guarded device. Its measurements are labeled as normal, indicating the legitimate state where the adversary has not yet tampered with the system. In each experiment we selected a group of other devices as rogue devices. Their measurements are labeled as anomalies, representing the state of the system after it was changed by an adversary.
  
  In all cases we used a train-test split of 60\%-40\% of the measurements labeled as normal. The measurements labeled as anomalies are not present in the training set.
  
  For the purpose of comparing the different feature sets, we set $T_{suspicion} = 1$. We then run our algorithm and calculate the false alarm and misdetection rates (FAR \& MDR respectively) as functions of $T_{votes}$. Next, we find the equal error rate (EER), the rate at which the FAR equals the MDR. The EER is the metric we use for comparing different hyper-parameters.
  
  In our graphs we convert the EER to ``false alarms per second'' (FA/Sec) under normal operation (system unaltered by an adversary). This gives a more concrete meaning to the numbers. The FA/Sec is calculated by multiplying the EER by the message rate, and is the inverse of mean time between failures. Note that each word occupies 36 bits, because the protocol mandates a minimum inter-word gap of at least 4 bit times. Thus the FA/Sec metric is defined as:
  
  \[FA/Sec = \frac{1}{MTBF} = EER \cdot \frac{100 {Kbits}/{sec}}{36bits}\]
  
  Note that since the FA/Sec is linear in the EER, we can discuss the graphs as though they display the EER when giving a qualitative analysis.
  
\sublevel{Identifying a Rogue Transmitter}
  In this series of experiments we simulate an attack, where the adversary switches the guarded transmitter LRU by a rogue transmitter LRU. In each experiment we designate one of the transmitters as the legitimate device to be guarded. In addition, we choose one receiver, either \(\text{P}_1\) or \(\text{P}_2\).  We train our IDS to identify samples from the chosen Tx-Rx pair as normal.
  
  We then test the trained IDS. We simulate a rogue transmitter LRU by using measurements of other transmitters connected to the chosen receiver as anomalies. We remind the reader that during each measurement, only one transmitter is connected to the bus.
  
  Only the Holt devices were used to simulate rogue transmitters, regardless of whether the guarded transmitter is an Excalibur (\(\text{E}_1\) or \(\text{E}_2\)) or a Holt (\(\text{H}_{10}\), ..., \(\text{H}_{13}\), \(\text{H}_{20}\), ..., \(\text{H}_{23}\)).
  
  For example, if we choose \(\text{E}_1\) as the guarded transmitter and \(\text{P}_1\) as the receiver, words from row 1 in Table \ref{tab:RecordingsSummery} are labeled as normal and used in the training stage and in the testing stage. Words from rows 6-13 are labeled as anomalies and used in the testing stage.
  
  We repeat this process for all possible values of $T_{votes}$ (0-127) while keeping $T_{suspicion} = 1$. For each value of $T_{votes}$ we indicate the MDR and the FAR. From these values we obtain the EER and the FA/sec.
  
  We repeat this process for four feature sets with all pairs of guarded transmitter and receiver. We end up with 18 experiments per feature set.
  
\begin{figure}[t]
    \centering
    \includegraphics[width=1.0\linewidth, angle=0]{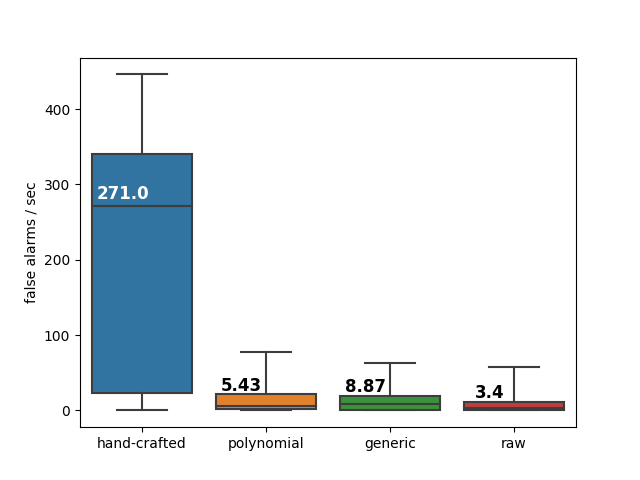}
    \caption{Comparing the feature sets for identifying a rogue transmitter.} 
    \label{fig:rogue_transmitter_results}
  \end{figure}

  The results are presented as a box plot in Figure \ref{fig:rogue_transmitter_results}. The x axis splits the results according to the used feature set. The y axis shows the false alarms per second, and 0 is the perfect score. The horizontal line in the middle of the box and the number written next to it indicate the median. The bottom and top boundaries of the box indicate the 1st and 3rd quartiles respectively. The horizontal lines outside of the box (the whiskers) indicate the minimum and maximum values.
  
  The figure shows that intruder detection yields the best results in term of EER when we use the Raw feature set. Both the median and the spread of the values are low. The EER values for the Generic and Polynomial feature sets are slightly more spread out, and the median is greater. The Hand-Crafted feature set is clearly inferior.
  
  The Generic, Raw and Polynomial feature sets have comparable performance, with Raw being slightly better with a median EER value of 0.12\% compared to 0.32\% and 0.19\% for the Generic and Polynomial feature sets. Since there is no significant reduction in memory costs from using the Generic feature set (recall Table \ref{tab:feature_set_sizes}), we conclude that in our case it is best to use the raw voltage samples, since in the trade-off between memory/runtime and performance, with the Generic set we spend significant effort to extract the features, and obtain no gain in comparison to the raw signal.
  
  We point out that there is a correlation between the number of features in the set and the performance of the feature set. The feature sets with reduced performance, namely the Hand-Crafted and Polynomial sets, have significantly fewer features for some segments - as few as 2 - and the Hand-Crafted sets totally ignores two segment types. The more features there are in the set, the more expressive the model is. Perhaps the two feature sets would perform better if they included additional features.
  
  Interestingly, for all feature sets there are experiments which reached a perfect EER value of 0. The guarded transmitters in these perfect experiments are \(\text{E}_1\), \(\text{E}_2\) and \(\text{H}_{10}\). Why do we achieve these results for \(\text{E}_1\) and \(\text{E}_2\)? We point out that we only use the Holt boards to simulate rogue devices. This means that in experiments where \(\text{E}_1\) and \(\text{E}_2\) are used as guarded devices, the IDS is tasked with differentiating between a guarded device and rogue devices that are manufactured by different vendors. We expect devices from different models to have different characteristics. However, we achieve \(\text{EER}=0\) for the Holt device \(\text{H}_{10}\) as a guarded device as well - indicating that even for the same manufacturer there are significant differences in the hardware fingerprint of individual devices.
  
\begin{figure}[t]
    \centering
    \includegraphics[width=1.0\linewidth, angle=0]{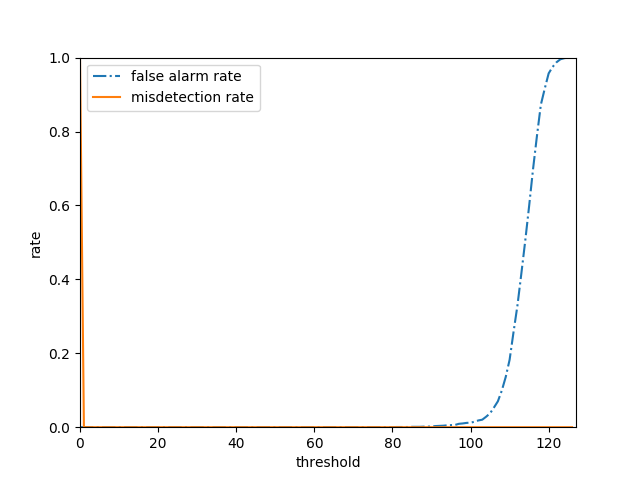}
    \caption{FAR and MDR for \(\text{E}_1\) as guarded as a function of the threshold}
    \label{fig:detection_easy_example}
  
    \includegraphics[width=1.0\linewidth, angle=0]{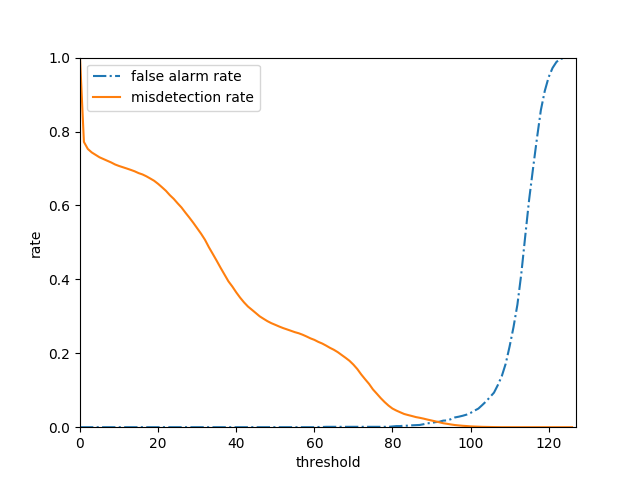}
    \caption{FAR and MDR for \(\text{H}_{21}\) as guarded as a function of the threshold}
    \label{fig:detection_difficult_example}
  \end{figure}

  We demonstrate this point by examining two selected experiments. We plot the MDR and the FAR as a function of the threshold value using \(\text{E}_1\) (Figure \ref{fig:detection_easy_example}) and of \(\text{H}_{21}\) (Figure \ref{fig:detection_difficult_example}) as guarded devices. Both are connected to \(\text{P}_1\) and in both figures the Raw feature set is used. Note that again, in these graphs, lower is better.
  
  Figures \ref{fig:detection_easy_example} and \ref{fig:detection_difficult_example} show that the two cases pose different levels of challenge for the IDS.
  In case of the \(\text{E}_1\) transmitter (Figure \ref{fig:detection_easy_example}), the MDR and the FAR curves do not intersect. In fact, the MDR overlaps the left-side boundary of the figure. There is a wide range of thresholds, for which an error rate of 0 can be achieved simultaneously for both rates. This makes \(\text{E}_1\) easily distinguishable from Holt transmitters. In contrast, in the case of the \(\text{H}_{21}\) transmitter (Figure \ref{fig:detection_difficult_example}) there is only a narrow range of thresholds for which both error rates are small, and the EER is greater than 0. This makes the tuning of the threshold important.
  
  Another thing to observe is that in both Figures \ref{fig:detection_easy_example} and \ref{fig:detection_difficult_example} the FAR curve is roughly the same, while the MDR curve spreads to higher thresholds in Figure \ref{fig:detection_difficult_example}. Note that the FAR is only calculated from samples of the guarded transmitter, and that the MDR is only calculated from samples of the rogue transmitters. The task of labeling a word from a guarded device as normal is not affected by the type of the guarded device. However, the success of the task of labeling rogue transmitters as anomalies heavily depends on the uniqueness of the guarded device.
  
\sublevel{Identifying a Rogue Receiver Switch}
  In the next series of experiments we simulate an attack, where the adversary replaces an existing receiver with a rogue one. The training stage is the same as the training stage in the previous series of experiments: One of the Holt devices is used as a transmitter, and either \(\text{P}_1\) or \(\text{P}_2\) is used a receiver.
  
  In the testing stage, we simulate an attack by using measurements taken with the same transmitter as in the training stage, but with the other receiver LRU.
  
  For example, if we choose \(\text{H}_{10}\) as the guarded transmitter and \(\text{P}_1\) as the receiver, words from row 6 in Table \ref{tab:RecordingsSummery} are labeled as normal and used in the training stage and in the testing stage. Words from rows 19 are labeled as anomalies and used in the testing stage. In other words, \(\text{P}_1\) as replaced by \(\text{P}_2\).
  
\begin{figure}[t]
    \centering
    \includegraphics[width=1.0\linewidth, angle=0]{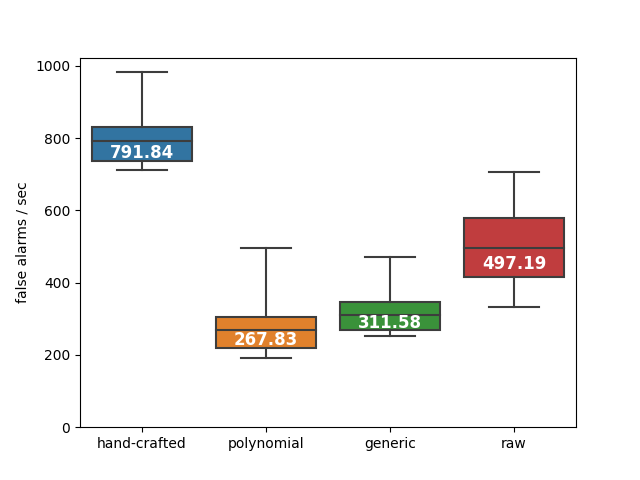}
    \caption{Comparing the feature sets for identifying a rogue receiver switch.}
    \label{fig:receiver_results}
  \end{figure}

  The results are shown in Figure \ref{fig:receiver_results}. We see that all feature sets perform worse than they did in identifying rogue transmitters (compare to Figure \ref{fig:rogue_transmitter_results}). That is to be expected, since communication systems are designed to mitigate the effect of the receivers on the transmission signal. The surprising result is that the Generic feature set performs better in this scenario than the Raw feature set, and the Polynomial feature set performs better than both of them. The fact that different feature sets are sensitive to changes in different components of the monitored system could be used to our advantage. An IDS could incorporate different anomaly detectors based on different feature sets. When an anomaly is detected, we could then identify the source of the attack by checking which feature set triggered the alarm.
  
\sublevel{Identifying an Addition of a Rogue Receiver}
  In the final series of experiments we simulate an attack, where the adversary connects an additional receiver to the bus. The training stage is the same as the training stage in the previous series of experiments, except that only \(\text{E}_1\) and \(\text{E}_2\) are used as transmitters.
  
  In the testing stage, we use measurements taken where a Holt receiver was connected to the bus in addition to the transmitter and receiver from the training stage.
  
  For example, if we choose \(\text{E}_1\) as the guarded transmitter and \(\text{P}_1\) as the receiver, words from row 1 in Table \ref{tab:RecordingsSummery} are labeled as normal and used in the training stage and in the testing stage. Words from rows 2-5 are labeled as anomalies and used in the testing stage.
  
\begin{figure}[t]
    \centering
    \includegraphics[width=1.0\linewidth, angle=0]{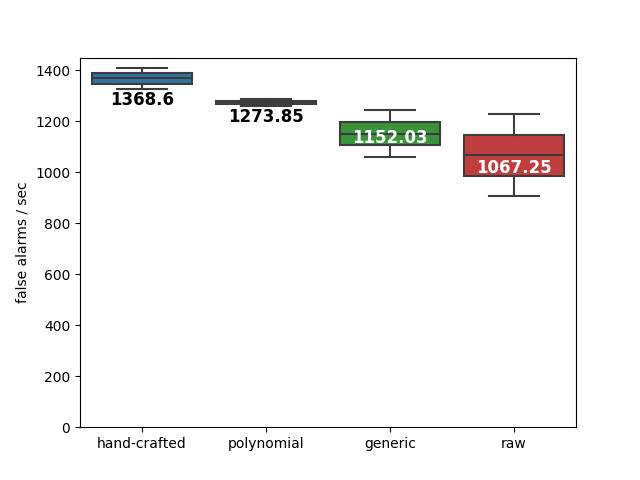}
    \caption{Comparing the feature sets for identifying a rogue receiver addition.}
    \label{fig:load_results}
  \end{figure}

  The results are shown in Figure \ref{fig:load_results}. Note that there are only two data points for each feature set (one for \(\text{E}_1\) and one for \(\text{E}_2\)). The numbers over the box plot represent the mean instead of the median.
  
  In this series of experiments all feature sets perform poorly. By comparing Figures \ref{fig:receiver_results} and \ref{fig:load_results} we learn that adding a receiver affects the electrical characteristics significantly less than replacing a receiver. This might stem from differences in hardware between the Holt evaluation boards and the proprietary receivers. Switching \(\text{P}_1\) for \(\text{P}_2\) (or vice versa) also means altering a long segment of the transmission line, while connecting an evaluation board receiver is accomplished by connecting two short wires to a header on our fabricated connector board. We interpret the results in the following way: The electrical characteristics are not especially sensitive to the receiver, but rather to the transmission line as a whole.
  
\begin{figure}
  \centering
  \begin{adjustbox}{width=\columnwidth}
  \begin{tikzpicture}[->,>=stealth',shorten >=1pt,auto,node distance=3cm,thick,
                      main node/.style={circle,draw,font=\sffamily\Large\bfseries},
                      skip node/.style={font=\sffamily\Large\bfseries}]
                      
    %
    
    \node[main node] (0) {0};
    \node[main node] (1) [right of=0] {1};
    \node[main node] (2) [right of=1] {2};
    \node[main node] (3) [right of=2] {3};
    
    \path[every node/.style={font=\sffamily\small}]
      (0) edge [out=210,in=150,looseness=10] node [left] {$1 - p$} (0)
          edge [bend left] node [above] {$p$} (1)
      (1) edge [bend left] node [below] {$1 - p$} (0)
          edge [bend left] node [above] {$p$} (2)
      (2) edge [bend left] node [below] {$1 - p$} (1)
          edge [bend left] node [above] {$p$} (3)
      (3) edge [out=30,in=330,looseness=10] node [right] {$1$} (3)
    ;
    
    %
    
  \end{tikzpicture}
  \end{adjustbox}
  \caption{Suspicion counter example for $T_{suspicion} = 3$}
  \label{fig:suspicion_counter}
  \end{figure}
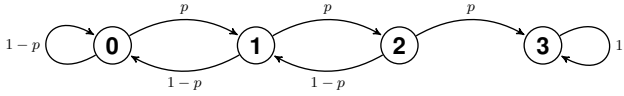

\level{Modeling the Suspicion Counter} \label{ModelingSuspicionCounter}
  In this section we analyze the effect of the suspicion counter on the overall false alarm rate, using a Markov-chain approach. Let the probability that a word is detected as anomalous be denoted by $p$, and assume that the events of detecting a word as anomalous are i.i.d. 
  Then we can describe the value of the counter after word $i$ arrives as a Markov random process. 
  Figure \ref{fig:suspicion_counter} shows, for example, the Markov process that corresponds to $T_{suspicion} = 3$.
  
  Each state in the diagram corresponds to a counter value. For a state representing a counter value of $0 < i < T_{suspicion}$ there is a transition to the right, with probability $p$, indicating that a word was detected as an anomaly (and the counter is increased), and a transition to the left, with probability $1-p$, indicating a normal word (and the counter is decreased). 
  The first and last states ($i = 0$ and $i = T_{suspicion}$ respectfully) are unique. The counter cannot be decreased below 0. Therefore, when $i = 0$, instead of a transition left we see a self-loop with probability $1 - p$. Finally, the state for $i = T_{suspicion}$ is a final state indicating that an alarm is raised: the probability of staying at $i = T_{suspicion}$ is 1. 
  
  The transition matrix of such a Markov process is given by:
  \begin{equation}
    P = \begin{bmatrix}
      1 - p & p     & 0      & 0 & \cdots & 0 \\
      1 - p & 0     & p      & 0 & \cdots & 0 \\
            &       & \vdots &   &       &   \\
      0     & \cdots & 0 & 1 - p  & 0 & p     \\
      0     & \cdots & 0 & 0      & 0 & 1
    \end{bmatrix}
  \end{equation}
  The vector \(x^{(n)}\) represents the distribution of probabilities between counter values after word $n$ was processed. The counter always starts at 0. Therefore:
  \begin{equation}
    x^{(0)} = \begin{bmatrix}
    1 & 0 & 0 & \cdots & 0
    \end{bmatrix}
  \end{equation}
  The distribution of probabilities after $n$ words is given by raising the transition matrix $P$ to the power of $n$:
  \begin{equation} \label{eq:prob_after_n_words}
    x^{(n)} = x^{(0)}P^n
  \end{equation}
   The probability of the counter reaching $T_{suspicion}$ after up to $n$ words is exactly the probability of reaching the final state $i = T_{suspicion}$, which is given by element $T_{suspicion} + 1$ of $x^{(n)}$.
   
  \begin{figure}[t]
    \centering
    \includegraphics[width=1.0\linewidth, angle=0]{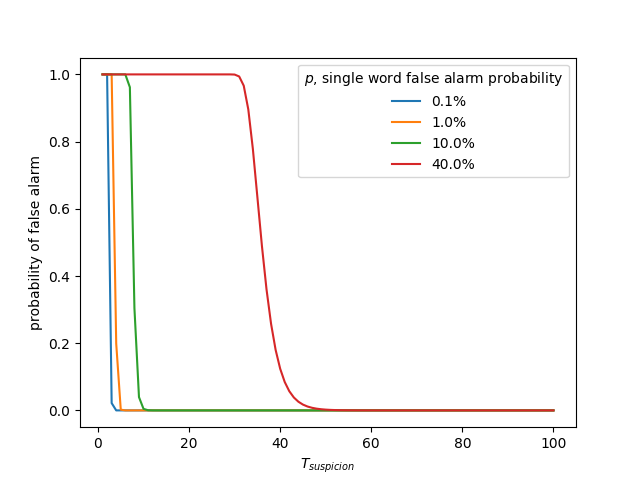}
    \caption{Probability for a false alarm during a 10-hour flight}
    \label{fig:CounterTheoreticalFA}
    
    \includegraphics[width=1.0\linewidth, angle=0]{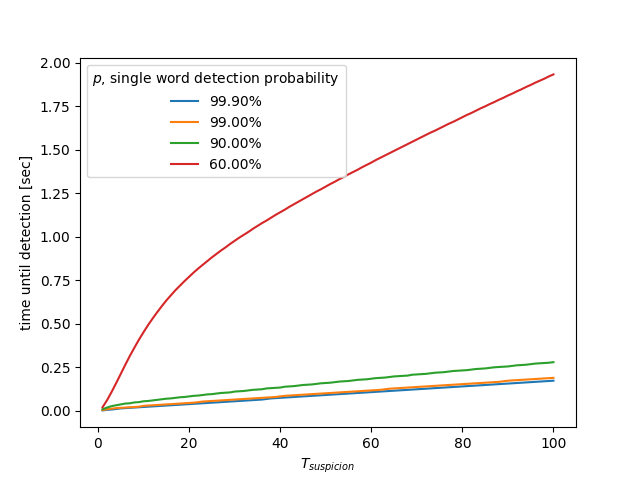}
    \caption{Time until the probability of a true detection exceeds 99.999\%}
    \label{fig:CounterTheoreticalTime}
  \end{figure}

  Figure \ref{fig:CounterTheoreticalFA} shows the probability of a false alarm occurring during a 10-hour flight as a function of $T_{suspicion}$ for different values of $p$. We assume an average transmission rate of 610 words per second, which is about 20\% of the maximal available bandwidth. This is the rate used in our data set. We can see that for every value of $p$, if $T_{suspicion}$ is high enough, the false alarm rate probability drops to 0. The lower $p$ is, the minimal $T_{suspicion}$ that is required is lower. Interestingly, even for a very high single-word false alarm probability of 40\%, at a $T_{suspicion}$ value of just 50 the probability of a false alarm drops to 0.
   
  Figure \ref{fig:CounterTheoreticalTime} show the time it takes for the probability for a positive (anomaly) detection to reach 99.999\%. Here all the transmitted words are assumed to originate from a rogue system, therefore we set $p > 0.5$. A low detection time means that the system is quick at detecting the adversary. The figure shows that the time until detection rises as $T_{suspicion}$ rises. The rise is quicker for low values of $p$. Even so, with a very poor detection probability of $p = 0.6$ and $T_{suspicion} = 100$, (which is much more than the threshold required to bring the false alarm rate to near 0), the time until detection reaches only 2 seconds. So, we can see from the Markov model that using a suspicion counter drastically reduces the false alarm rate, while slowing down the detection only mildly.
   
\level{Performance of the Complete Method} \label{PerformanceEvaluationCompleteMethod}

  The results we attained in Section \ref{PerformanceEvaluationSingleWord} are encouraging: we can successfully fingerprint a transmitter based on a single word. However, the FA/Sec metric of around 5 alarms per second is still too high for a realistic system. To reduce the FA/Sec rate to near-zero, we use the suspicion counter we analyzed in Section~\ref{ModelingSuspicionCounter}, and raise an alarm only when the counter exceeds $T_{suspicion}$. In this section we empirically evaluate the behavior of the complete system as a function of the threshold $T_{suspicion}$.
  
We do not discuss how to identify an additional receiver, since we could not identify it with sufficient certainty in Section \ref{PerformanceEvaluationSingleWord}.
  
  In Section \ref{PerformanceEvaluationSingleWord} we saw that different feature sets are suited for detecting different adversary models; the Raw feature set for detecting a rogue transmitter, and the Polynomial feature set for detecting a rogue receiver. We now continue our evaluation with these two feature sets.
  
\begin{figure}[t]
    \centering
    \includegraphics[width=1.0\linewidth, angle=0]{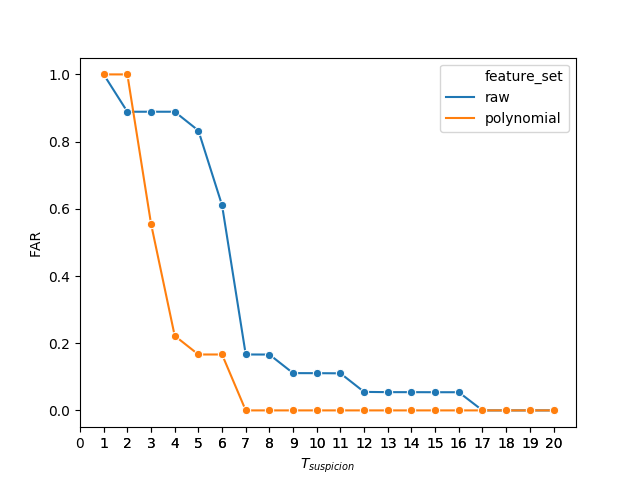}
    \caption{FAR as a function of $T_{suspicion}$. $T_{votes} = 100$}
    \label{fig:counter_far}
  \end{figure}

  We wish to examine the FAR as a function of the counter threshold. For each transmitter-receiver pair in our data set we repeat the following procedure. First, we train an anomaly detector on words recorded with the selected pair. Then we test the detector 1000 times on words recorded with the same pair. Each time we use the same 1968 words after cyclically shifting them by a random integer in order to start the counter at a different point in time. We compute the FAR by dividing the number of times an anomaly was detected by the total number of measurements. Overall there are such 18000 measurements for each data set (9 transmitters $\times$ 2 receivers $\times$ 1000 repetitions).
  We repeat this procedure with different values of $T_{suspicion}$, once for each feature set. We use $T_{votes} = 100$ in all experiments. Experiments in Section \ref{PerformanceEvaluationSingleWord} show that this is a reasonable choice that balances the FAR and the MDR for detection based on single words. The train-test split is 60\%-40\%. Figure \ref{fig:counter_far} shows the results.
  
  As predicted by the Markov analysis, the false alarm rate drops dramatically as $T_{suspicion}$ increases. For $T_{suspicion}$ greater than 16, there were no false alarms: the empirical results and the Markov analysis are in agreement, and the empirical Figure \ref{fig:counter_far} is similar to the theoretical Figure \ref{fig:CounterTheoreticalFA}. In both the false alarm probability starts at 1, is stable for low values of $T_{suspicion}$, drops quickly and finally decays to 0.
  
  For observing the trade-off of using an anomaly counter, measuring the MDR rate is inefficient, since given sufficient time an anomaly will eventually be detected. Instead of measuring the MDR, we measure the \emph{time} it takes for our detector to raise an alarm. The procedure is similar to the procedure of measuring the FAR. Instead of testing the trained detector on words recorded with the same transmitter-receiver pair as the one on which it was trained, we test it on words from other pairs, according to the adversary model that is being simulated, as explained in section \ref{Methodology}. The Raw feature set is used for measuring a rogue transmitter detection, and the polynomial feature set is used when detecting a rogue receiver. We count the number of words it takes for the detector to raise an alarm. Overall, for a rogue transmitter there are 144000 measurements (9 guarded transmitters $\times$ 8 rogue transmitters $\times$ 2 receivers $\times$ 1000 repetitions) and for a rogue receiver there are 8000 measurements (8 transmitters $\times$ 1 guarded receiver $\times$ 1 rogue receiver $\times$ 1000 repetitions) for each value of $T_{suspicion}$.
  
\begin{figure}[t]
    \centering
    \includegraphics[width=1.0\linewidth, angle=0]{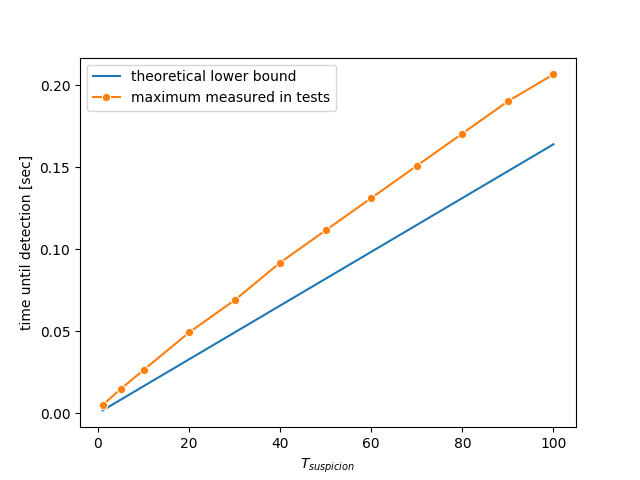}
    \caption{Maximal length of time for detecting a rogue transmitter. $T_{votes} = 100$}
    \label{fig:counter_detection_time_rogue_tx}
  
    \includegraphics[width=1.0\linewidth, angle=0]{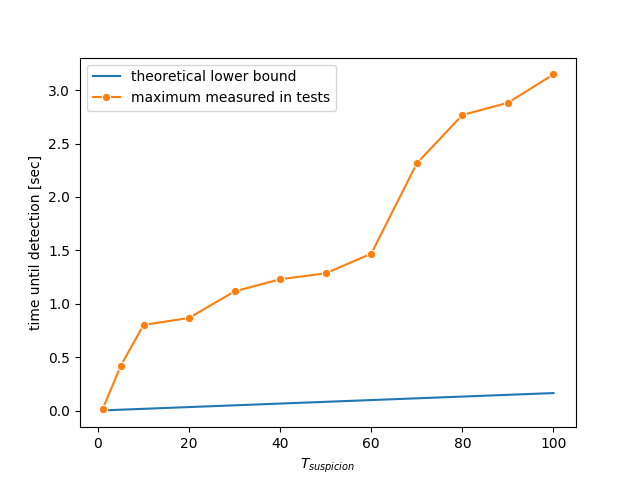}
    \caption{Maximal length of time for detecting a rogue receiver. $T_{votes} = 100$}
    \label{fig:counter_detection_time_rogue_rx}
  \end{figure}

  Figure \ref{fig:counter_detection_time_rogue_tx} shows the maximal (worst case) time we measured for detecting a rogue transmitter over all combinations of guarded and attacking transmitters. Figure \ref{fig:counter_detection_time_rogue_rx} shows the same for detecting a rogue receiver. The blue line indicates a lower bound---the time it takes to transmit $T_{suspicion}$ messages. In our test set the average transmission rate is 610 words per second, which is about 20\% of the maximal available bandwidth.
  
  The figures show that the suspicion counter reduces false alarms at the expense of only mildly delaying the detection of an attack. We find the trade-off worthwhile. Even when $T_{suspicion} = 20$ and there are no false alarms, a rogue transmitter is detected in under 50ms, which greatly reduces the adversaries ability to mount a successful attack, and a rogue receiver is detected in several seconds, which is still good.
  
  In both cases, the maximal observed detection time rises faster than lower bound. For rogue transmitter detection, the maximal time is of the same order of magnitude of as the lower bound, while for the case of rogue receiver detection, the maximal time is an order of magnitude higher. The gap is explained by the difference in misdetection rates for single words, measured in section \ref{PerformanceEvaluationSingleWord}. The difference between the slopes is predicted by the Markov analysis. Figure \ref{fig:CounterTheoreticalTime} shows that as the misdetection rate of a single word increases, so does the rate at which the detection time rises.
  
\level{Conclusions} \label{Conclusions}
  We presented a first hardware fingerprinting method for the ARINC 429 bus. The method can be used to retrofit source authentication into existing avionic systems with low effort.
  
  We showed that our method is especially effective for identifying a technician attack, in which an adversary replaces a legitimate LRU with a rogue one. We demonstrated that even transmitter LRUs of the same make and model are different enough from one another for them to generate distinguishable signals. All the more so when dealing with devices from different vendors. We found that skipping the feature extraction stage and using the raw signal achieves the best result.
  
  We showed that the method can also detect a switched receiver. We found that the Polynomial feature set, which was conceived for the purpose of this \iftoggle{paper} {paper} {work}, achieves the best performance among the feature sets we examined, when applied to this task.

  We showed that by augmenting the per-word anomaly detection by a ``suspicion counter'', we can drastically reduce the false-alarm rate. Using both a Markov-chain analysis and an extensive empirical evaluation, we demonstrated that our full intrusion detection system is quite realistic: e.g., it achieves near-zero false alarms per second, while detecting a rogue transmitter in under 50ms, and detecting a rogue receiver in under 3 seconds. In other words, technician attacks can be reliably detected during the pre-flight checks, well before the aircraft takes off.
  
  Further work needs to be done in order to evaluate the sensitivity of the hardware fingerprints to external changes such as fluctuations in temperature or supply voltage levels, and to evaluate its stability over time.
  
  ARINC 429 lacks essential security features. It is a safety liability that is present today in almost every civil aircraft.
  Our method could help close of the gap between ARINC 429 and modern security requirements. Thus, we argue that it is a valuable addition to the protection of any airborne system which uses the ARINC 429 bus.
  
\bibliographystyle{IEEEtranS}


\iftoggle{paper}{
}{
  \pagenumbering{gobble}  
  \pagestyle{empty}
  \newchapterevenpage
  \input{abstract_hebrew.tex}
  \newchapterevenpage
  \input{title_page_hebrew_inner.tex}
  \newchapterevenpage
  \input{title_page_hebrew_outer.tex}
}

\end{document}